\documentclass[twocolumn,twoside,slac]{revtex4}
\usepackage{graphicx}
\usepackage{fancyhdr}
\usepackage{amssymb}
\newcommand{\df}{\,\mathrm{d}}
\begin{document}
\title{
A comparison of methods for confidence intervals
}

\author{ A.Bukin} %
\affiliation{Budker INP, 630090 Novosibirsk, Russia}

\begin{abstract}
Comparisons are carried out of the confidence intervals
constructed with Neyman's frequentist method and with
the $\Delta L=1/2$ likelihood method, using the example of
low-statistics life time estimates.
\end{abstract}

\maketitle
\thispagestyle{fancy}


\section{%
P.d.f. for life time estimators%
}

For a given value $\tau$ of the true life time, the P.D.F. of
a measurement is
$$
\frac{\df W}{\df t}=\frac{1}{\tau}\exp\left(-\frac{t}{\tau}\right),$$
and so for an experiment with $n$ measurements
\begin{equation}\label{eq:Mutual_probability}
\df W=\frac{1}{\tau^n}\cdot\prod\limits_{k=1}^n\df t_k\cdot\exp\left(
-\frac{t_k}{\tau}\right).
\end{equation}
The negative log likelihood function is
\begin{equation}
L=n\ln\tau+\frac{1}{\tau}\cdot
\sum\limits_{k=1}^nt_k.
\end{equation}
The maximum likelihood estimator of the lifetime can easily be
found minimizing $L$
\begin{equation}
\hat{\tau}=\frac{1}{n}\sum\limits_{k=1}^nt_k;\;\;\;
\min L =L_0=n+n\ln\hat{\tau},
\end{equation}
so the probability~(\ref{eq:Mutual_probability})
can be transformed to
\begin{equation}
\frac{\df W}{\df\hat{\tau}}=\frac{1}{(n-1)!}\cdot
\left(\frac{n\hat{\tau}}{\tau}\right)^{n-1}\cdot\frac{n}{\tau}
\cdot\exp\left(-\frac{n\hat{\tau}}{\tau}\right).
\end{equation}

Given some true value $\tau$ then for any algorithm that defines
a confidence interval
 $\hat{\tau}^{+\Delta\tau^{(+)}}%
_{-\Delta\tau^{(-)}}$ we can
evaluate the coverage  $P$:
\begin{equation}\label{eq:Coverage_definition}
P=\frac{1}{(n-1)!}\cdot\frac{n}{\tau}\cdot
\int\limits_{\hat{\tau_1}}^{\hat{\tau_2}}
\left(\frac{n\hat{\tau}}{\tau}\right)^{n-1}\cdot
\exp\left(-\frac{n\hat{\tau}}{\tau}\right)\df\hat{\tau},
\end{equation}
where $\hat{\tau_1}+\Delta\tau^{(+)}=\tau;\;\;\;
\hat{\tau_2}-\Delta\tau^{(-)}=\tau.$

\section{%
 Likelihood Function confidence interval%
}

The conventional Likelihood function method for finding
a 68\% confidence interval
\cite{Hudson,LF_method1} is to find the values of $\tau$ for which
$$\Delta L=L-L_0=\frac{1}{2}. $$
In our case
\begin{equation}
\Delta
L=n\cdot\left(\frac{\hat{\tau}}{\tau}-1+
\ln\frac{\tau}{\hat{\tau}}\right)=\frac{1}{2}.
\end{equation}
For example, for $n=5$ the limits are
\begin{equation}\label{LoglikelihoodCI}
\left[\tau_1,\tau_2\right]=\left[0.6595\hat{\tau},1.6212\hat{\tau}\right].
\end{equation}
 The coverage of this interval, from
 Equation~(\ref{eq:Coverage_definition}), is
$$
\frac{1}{4!}\cdot\frac{5}{\tau}\cdot
\int\limits_{\hat{\tau_1}}^{\hat{\tau_2}}
\left(\frac{5\hat{\tau}}{\tau}\right)^4\cdot
\exp\left(-\frac{5\hat{\tau}}{\tau}\right)\df\hat{\tau}=0.6747,\;\;$$
where the integration limits,
corresponding to~(\ref{LoglikelihoodCI}), are
$$\left[\hat{\tau_1},
 \hat{\tau_2}\right]=\left[0.6168\tau,1.5163\tau\right].$$
{The coverage is close to, but significally
different from, the nominal value of  0.6827.}

\begin{figure}
\includegraphics[width=0.4\textwidth]{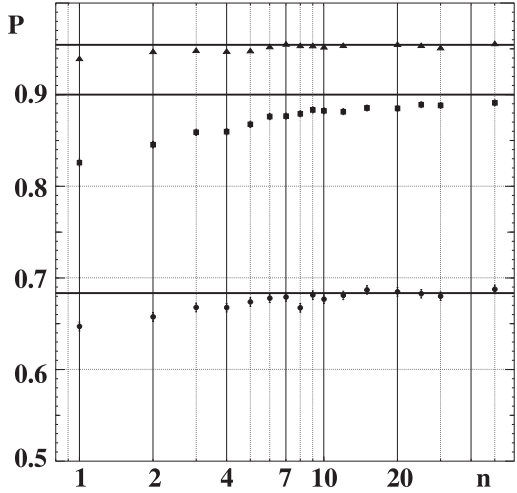}
\caption{\label{LF_coverage}Coverage for likelihood function
confidence intervals, evaluated by Monte Carlo. Statistical
errors are shown when they exceed the size of polymarker.
$\blacktriangle$ --- 95\% Conf.Interv.,
$\blacksquare$ --- 90\% Upper limit,
$\bullet$ --- 68\% Conf.Interv.}
\end{figure}

Examples of
confidence intervals obtained by this means are shown
 in Table~\ref{Mutual_Neyman_LF_tabl}, as the values in parantheses.
The 95\% confidence interval  was obtained using the
 rule $\Delta L=2$, and 90\% upper limit using a one side interval
for which
$$
\Delta L=\left[\mbox{erf}^{-1}\left(2\cdot 0.9-1\right)\right]^2
\approx 0.821.$$
The  coverage given by such intervals is shown
in Fig.\,\ref{LF_coverage},
evaluated using a Monte Carlo method.

\section{%
 Bayesian confidence interval%
}
For comparison we can estimate a Bayesian confidence interval
for the same example of $n=5$.
In the Bayesian approach \cite{Jeffreys,BayesAll,Neyman2},
 the likelihood
function is considered to be a probability density for the true
 parameter $\tau$. Assuming a flat prior distribution for $\tau$
 this is
$$
\frac{\df W}{\df\tau}=P(\tau)\sim{\cal L}=
\frac{1}{\tau^n}\cdot \exp\left(-\frac{n\hat{\tau}}{\tau}\right).
$$
After normalization (for $n\geq 2$) this becomes:
$$
P(\tau)
=\frac{\left(n\hat{\tau}\right)^{n-1}}
{(n-2)!\cdot \tau^n}\cdot\exp\left(-\frac{n\hat{\tau}}{\tau}\right),
$$
which for $n=5$ gives
\begin{equation}\label{BayesianPDF}
\int\limits_0^\tau\df W=\left[1+\frac{5\hat{\tau}}{\tau}+
\frac{1}{2}\left(\frac{5\hat{\tau}}{\tau}\right)^2+
\frac{1}{6}\left(\frac{5\hat{\tau}}{\tau}\right)^3\right]\cdot
e^{-\frac{5\hat{\tau}}{\tau}}.
\end{equation}

The 68\% central confidence region for this distribution is
(see Fig.\,\ref{Bayes_distribution}):
\begin{figure}[tbp]
\centerline{\includegraphics[width=0.4\textwidth]{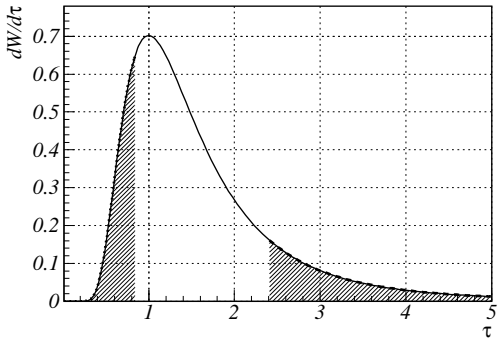}}
\caption{\label{Bayes_distribution}Probability density function
for the true value of the parameter $\tau$ in a Bayesian approach
(Equation~(\ref{BayesianPDF}) with $n=5$ and $\hat{\tau}=1$).
The shaded regions are the 16\% ``tails''.}
\end{figure}
$$
\tau=\hat{\tau}\cdot\left(1^{+1.3974}_{-0.1552}\right)\;\;
$$
The coverage of this region is actually not 68.27\% but 64.31\%.

\section{%
 Neyman's confidence interval%
}

\begin{figure}[tbp]
\begin{center}
\begin{minipage}[b]{0.35\textwidth}
\includegraphics[width=\textwidth]{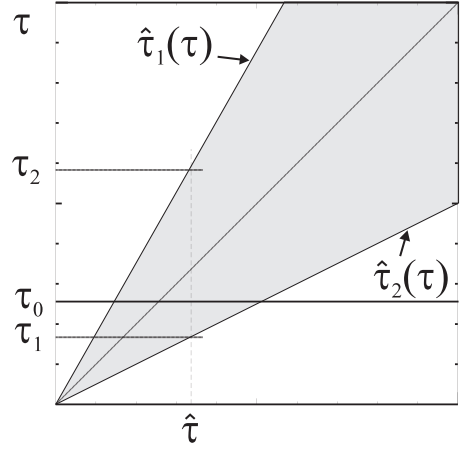}
\end{minipage}
\end{center}
\caption{\label{CZ_illustration}Illustration of the construction
of a confidence zone
(or confidence belt).}
\end{figure}

Neyman~\cite{Neyman1,Neyman2,Barlow}
proposed a frequentist construction of a  confidence zone
(or confidence belt) as follows (see Figure~\ref{CZ_illustration}):
\begin{enumerate}
\item One obtains functions $\hat{\tau}_1(\tau)$ and
$\hat{\tau}_2(\tau)$ of the true parameter $\tau$
  such that
$$
\begin{array}{l}
\int\limits_{0}^{\hat{\tau}_1(\tau)}
\frac{\df W(\hat{\tau};\tau)}{\df \hat{\tau}}\df \hat{\tau}=\frac{1-\beta}{2};
\;\;\;\\[4mm]
\int\limits_{\hat{\tau}_2(\tau)}^\infty
\frac{\df W(\hat{\tau};\tau)}{\df \hat{\tau}}\df \hat{\tau}=\frac{1-\beta}{2},
\;\;\;
\end{array}
$$
where $\beta$ is the confidence level required, here $\beta=0.6827$.
For  $n=5$ these are simply  $\hat{\tau}_1(\tau)=0.568\tau$,
$\hat{\tau}_2(\tau)=1.433\tau$, as shown in Figure~\ref{CZ_illustration}.
\item One defines the inverse functions
$$\tau_1(\hat{\tau})=\hat{\tau}_2^{-1}(\hat{\tau});\;\;\;
\tau_2(\hat{\tau})=\hat{\tau}_1^{-1}(\hat{\tau})
$$
which, for a given value of $\hat{\tau}$,
define the borders of the confidence interval for $\tau$,
with
coverage $\beta$.

In our example, there are
 $\tau_1(\hat{\tau})=0.698\hat{\tau}$,
$\tau_2(\hat{\tau})=1.760\hat{\tau}$.

Thus the
result of a lifetime experiment of this type can be written
$$\tau=\hat{\tau}\cdot\left(1^{+0.760}_{-0.302}
\right).$$

The coverage evaluated is 0.6826 --- the difference of 0.0001
is purely due to
rounding errors.
\end{enumerate}
Table~\ref{Mutual_Neyman_LF_tabl} shows these  intervals
for several values of $n$,
with the likelihood approximation shown in  parantheses
for comparison.

Table~\ref{Three_methods} compares the coverage of all three
 methods for the  $n=5$ case.

\begin{table*}[tbp]
\caption{\label{Mutual_Neyman_LF_tabl}Lifetime
confidence intervals
obtained by Neyman's method for various values of $n$,
the number of measurements, and confidence levels.
}
\begin{center}
\begin{tabular}{|c|c|c||c|c||c|}
\hline
\raisebox{-3mm}[0mm][0mm]{$n$} &
\multicolumn{2}{|c||}{68\% C.L.} &
\multicolumn{2}{|c||}{95\% C.L.} & 90\% C.L.\\
\cline{2-5}
 & $\frac{\Delta\tau^{(-)}}{\hat{\tau}}$ &
$\frac{\Delta\tau^{(+)}}{\hat{\tau}}$&
 $\frac{\Delta\tau^{(-)}}{\hat{\tau}}$ &
 $\frac{\Delta\tau^{(+)}}{\hat{\tau}}$ & upper limit\\
\hline
1 &0.457 (0.576) &4.789 (2.314)  & 0.736 (0.778) &42.45 (18.06)&
  $9.49\hat{\tau}$ ($8.49\hat{\tau}$)\\
2 &0.394 (0.469) &1.824 (1.228)  & 0.648 (0.682) & 7.690 (5.305)&
  $3.76\hat{\tau}$ ($2.76\hat{\tau}$)\\
3 & 0.353 (0.410) &1.194 (0.894)  & 0.592 (0.621) & 4.031 (3.164) & \\
4 & 0.324 (0.370) &0.918 (0.725)  & 0.551 (0.576) & 2.781 (2.314) & \\
5 & 0.302 (0.341) &0.760 (0.621)  & 0.519 (0.541) & 2.159 (1.858) & \\
6 & 0.284 (0.318) &0.657 (0.550)  & 0.492 (0.513) & 1.786 (1.571) & \\
7 & 0.270 (0.299) &0.584 (0.497)  & 0.470 (0.489) & 1.538 (1.374) & \\
8 & 0.257 (0.284) &0.529 (0.456)  & 0.452 (0.469) & 1.359 (1.228) & \\
9 & 0.247 (0.271) &0.486 (0.423)  & 0.435 (0.451) & 1.225 (1.116) & \\
10& 0.237 (0.260) &0.451 (0.396)  & 0.421 (0.436) & 1.119 (1.027) & \\
20& 0.182 (0.194) &0.285 (0.261)  & 0.331 (0.341) & 0.654 (0.621) & \\
50& 0.124 (0.129) &0.164 (0.156)  & 0.232 (0.237) & 0.356 (0.346) &\\
\hline
\end{tabular}
\end{center}
\end{table*}

\begin{table}[tbp]
\caption{\label{Three_methods}Coverage of all three methods for $n=5$}
\begin{center}
\begin{tabular}{|l|c|c|c|}
\hline
Method & Negative error & Positive error & Coverage, \% \\
   & ${\Delta\tau^{(-)}}/{\hat{\tau}}$ &
     ${\Delta\tau^{(+)}}/{\hat{\tau}}$ & \\
\hline
Likelihood & 0.341 & 0.621 & 67.47 \\
Bayesian   & 0.155 & 1.397 & 64.31 \\
Neyman's   & 0.302 & 0.760 & 68.26 \\
\hline
\end{tabular}
\end{center}
\end{table}

\vspace{30mm}

\section{
Conclusion
 }

\begin{itemize}
\item
Neyman's method for  confidence
intervals provides exact coverage, by construction.
\item
The intervals from $\Delta L=1/2$ agree well with the Neyman intervals
for large $n$, but differ for small $n$, as can be seen
in Table~\ref{Mutual_Neyman_LF_tabl}.
In such cases they undercover, i.e. the interval is smaller than the
true one.
\item
Bayesian confidence intervals give
very different results, and can undercover or overcover.
\end{itemize}

\end{document}